\documentclass{appolb}
\usepackage{graphicx}
\usepackage{amsmath}
\usepackage{cmap}
\usepackage{slashed}

\begin{document}
\title{Decay mechanisms in bound state interaction kernels%
\thanks{Presented by A.S. Miramontes at ``Excited QCD 2018'', Kopaonik, Serbia, March 11-15 2018  }%
}
\author{A.S. Miramontes, H. Sanchis-Alepuz
\address{Institut f{\"u}r Physik, Karl-Franzens-Universit\"at Graz, NAWI Graz, 8010 Graz, Austria}
\\
}
\maketitle
\begin{abstract}
We present progress on the study of decay-channel effects in the properties of hadrons using covariant Bethe-Salpeter equations (BSEs). The main goal will be to develop BSE kernels that contain explicit decay mechanisms. This will be first explored in the meson sector where, for example, a kernel suitable to study the rho meson should contain a virtual  $\rho \rightarrow  \pi + \pi$ decay mechanism.​ This will be tackled by including explicit pion degrees of freedom in addition to quarks and gluons.
\end{abstract}
  
\section{Introduction}


Dyson-Schwinger (DSE) and Bethe-Salpeter (BSE) equations are extensively used to study hadron phenomenology (see, e.g. \cite{Eichmann:2016yit} and references therein). In practice, truncations of the DSEs and BSEs are necessary. The simplest one which preserves chiral symmetry in the chiral limit is the rainbow-ladder (RL) truncation, and it has been successfully used for ground state meson and baryon calculations \cite{Eichmann:2016yit, Eichmann:2013afa}. In this truncation, hadrons are stable bound states and they do not decay. Nevertheless, most hadrons are resonances and they do decay. In order to get a complete description of hadrons we must incorporate these features. 

\section{Formalism}

\subsection{Dyson-Schwinger equations}
\begin{figure}[t]
\centerline{%
\includegraphics[width=10.5cm]{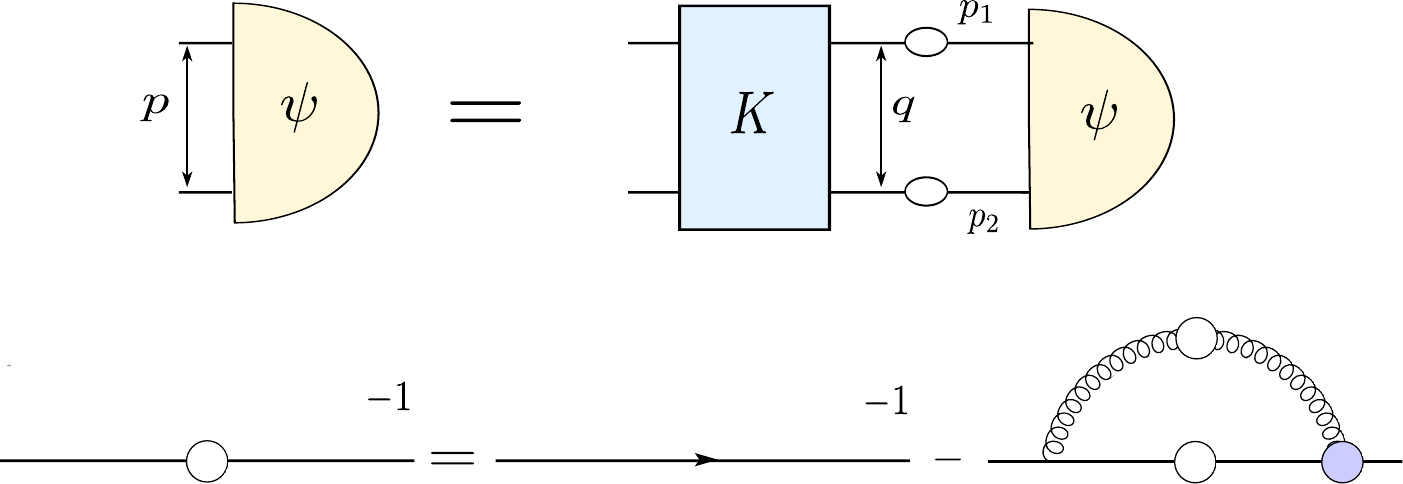}}
\caption{Diagrammatic representation of the two body Bethe Salpeter equation (upper) and the Dyson-Schwinger equation for the quark propagator (lower).}
\label{Fig:DSE}
\end{figure}

The fully-dressed quark propagator can be obtained as a solution of a DSE, given diagrammatically in Figure \ref{Fig:DSE}. The fully-dressed inverse quark and bare propagator are parametrized as
\begin{equation}
S^{-1}(p) = i \slashed{p} A(p^2) + B(p^2)~, \qquad  S_0^{-1}(p) = i \slashed{p} + m~.
\end{equation}

\noindent Then, the quark DSE is given by,
\begin{equation}
S^{-1} = Z_2  S_0^{-1} - Z_{1f} \int \frac{dq^2}{(2 \pi)^4} \gamma_\mu S(q) \Gamma_\nu(q,k) D_{\mu \nu}(k)~,
\end{equation}

\noindent where $Z_{1f}$ and $Z_2$ are renormalization constants. The quark mass functions $M(p^2)$ is given by $M(p^2) = B(p^2)/A(p^2)$ and $D_{\mu \nu}(k)$ is the full gluon propagator.
\begin{equation}
D_{\mu \nu}(k) = \left( \delta_{\mu \nu} - \frac{k_\mu k_\nu}{k^2}\right) \frac{Z(k^2)}{k^2}
\end{equation}

As already mentioned, in practical calculations the quark DSE has to be truncated. In the RL truncation of the quark DSE we keep only the vector part of the full quark-gluon vertex $\Gamma$ and collect all the dressings into an effective coupling $\alpha(k^2)$, namely $\gamma_\mu Z(k^2) \Gamma_\nu(q,k) \rightarrow \gamma_\mu \alpha(k^2) \gamma_\nu$.
To describe the coupling $\alpha(k^2)$ one needs to use a model. A very popular choice is that of Maris and Tandy \cite{Maris:1997tm, Maris:1999nt}.
%
%

%


\subsection{Bethe-Salpeter equations} 
\begin{figure}[t]
\centerline{%
\includegraphics[width=10.5cm]{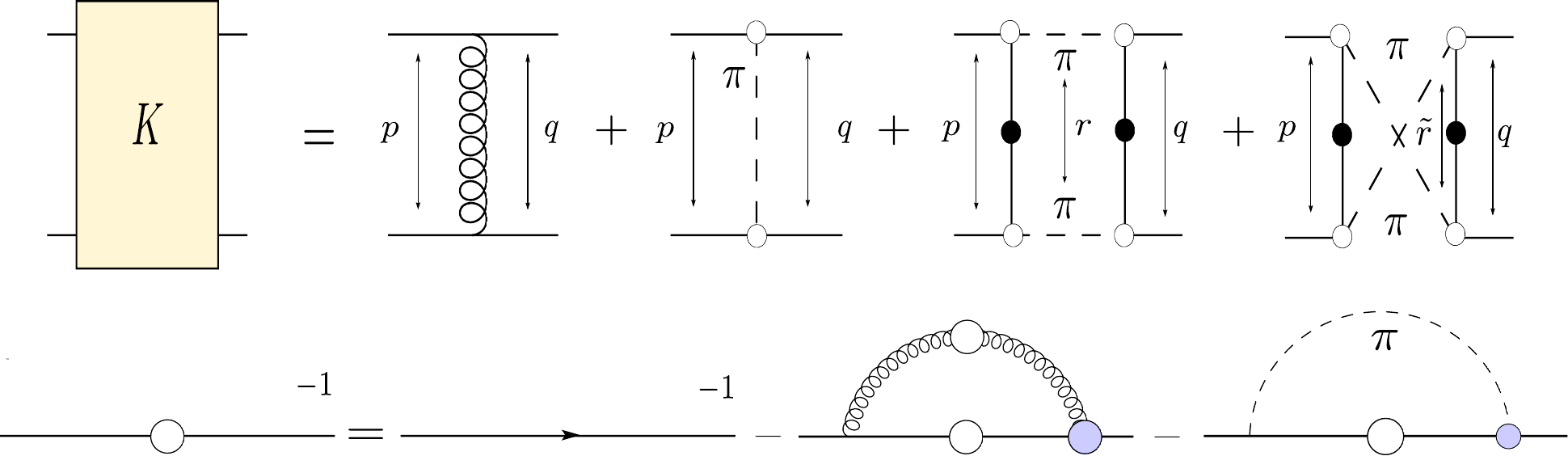}}
\caption{Interaction kernel including t, s and u-channels (upper), and the Dyson-Schwinger equations for the quark propagator with one gluon exchange and one pion exchange (lower), where $p,q,r$ and $\tilde{r}$ are the quark relative momentum. }
\label{Fig:newkernel}
\end{figure}

If the n-particle system forms a bound state, a pole appears in the Green function for $P^2 = -M^2 + iM\Gamma$,
\begin{equation}
 G \rightarrow \frac{\Psi \bar{\Psi}}{P^2 + M^2 -iM\Gamma},\label{eq:pole}
\end{equation}

\noindent where $\Psi$ defines the Bethe-Salpeter amplitude, $M$ is the mass of the bound state and $\Gamma$ its width. From Eq.~\eqref{eq:pole} a homogeneous equation for $\Psi$ can be obtained,
\begin{equation}
\Psi(p;P) = \int \frac{d^4 q}{(2\pi)^4}K(p,q;P) S\left(p + \frac{P}{2}\right) \Psi(q;P)S\left(q - \frac{P}{2}\right),
\label{eq.BSE}
\end{equation}

\noindent with $q$ the quark relative momentum. Equation~\eqref{eq.BSE} is known as the Bethe-Salpeter equation. As with the quark DSE, the BSE interaction kernel $K$ must be truncated. In the RL truncation, the truncated kernel corresponds to a single gluon exchange (see Fig.~\ref{Fig:newkernel} first term).


%

%

%

\section{Beyond rainbow-ladder}
In RL truncation the hadrons are stable particles. As such, they appear as poles on the negative real axis of the complex $P^2$ plane (cf. Eq.~(\ref{eq:pole}). In order to shift the pole away from the real axis and turn the state into a resonance with a non-vanishing width, the interaction kernel $K$ must allow for \textit{virtual} decays into suitable channels. As a workbench, we will study the rho meson and consider the width corresponding to the $\rho\rightarrow\pi\pi$ strong decay\footnote{During the preparation of this work an analogous calculation has appeared \cite{Williams:2018adr}.}. In \cite{Fischer:2007ze,Fischer:2008sp,Sanchis-Alepuz:2014wea}, pionic effects where considered by including explicit pionic degrees of freedom in the system, in addition to quarks and gluons. Thereby, besides the gluon part of the quark DSE, an emission and absorption of a pion appears.

\subsection{The t-channel pion exchange}
The resulting DSE for the quark propagator is diagrammatically represented in Fig.~\ref{Fig:newkernel} and is given by \cite{Fischer:2007ze,Fischer:2008sp},
\small
\begin{eqnarray}
S^{-1}(p) &=& S^{-1}(p)^{RL} - 3 \int \frac{d^4 q}{(2\pi)^4}\Bigg[Z_2 \gamma_5 S(q)\Gamma_{\pi}\left(\frac{p+q}{2}, q-p\right) \nonumber \\
&+& Z_2 \gamma_5S(q)\Gamma_{\pi}\left(\frac{p+q}{2}, p-q\right)\Bigg] \frac{D_{\pi}(k)}{2},
\end{eqnarray}
\normalsize
\noindent where the pion propagator is given by $D_{\pi}(k) = (k^2 + m_{\pi}^2)^{-1}$. The pion-quark vertices are given by the pion Bethe Salpeter amplitude, which can be represented as 
\small
\begin{equation}
\Gamma^i_{\pi}(p,P) = \tau^i \gamma_5 \left\lbrace E_{\pi}(p,P) - i \slashed{P} F_{\pi}(p,P) - i\slashed{p} (p \cdot P) G_{\pi}(p,P) - \left[ \slashed{P},\slashed{p}\right]H_{\pi}(p,P) \right\rbrace ~,
\end{equation}
\normalsize
with $E_{\pi}, F_{\pi} , G_{\pi}, H_{\pi}$ four independent dressing functions. 

\noindent In addition, the Bethe-Salpeter kernel is constrained via the axial Ward-Takahashi identity (axWTI) in order to fulfil the Goldstone boson property of the pion in the chiral limit,
\small
\begin{equation}
\left[\Sigma(p_+) \gamma_5 + \gamma_5\Sigma(p_-)\right]_{tu} = \int \frac{d^4 k}{(2 \pi)^4} K_{tu;sr}(p,k;P)\left[\gamma_5 S(k_-) + S(k_+)\gamma_5 \right]_{rs} 
\end{equation}
\normalsize
\noindent In ref.~\cite{Fischer:2007ze} the authors considered only the t-channel contribution to $K$, namely, the second term in the upper panel of Fig.~\ref{Fig:newkernel}. However, they noticed that a straightforward implementation of that diagram violated the axWTI, but they found a particular combination of terms that restores the symmetry.

\subsection{The s- and u-channel pion exchange}

The decay mechanism corresponds to the s- and u-channels (resp. third and fourth terms) in Fig.~\ref{Fig:newkernel}. As it was done in \cite{Fischer:2007ze,Fischer:2008sp} we might first need to modify their structure in order to preserve the relevant symmetries. In this case, however, we cannot use the axWTI to fix it because the contributions of s- and u-channels are identically zero. The kernels for the s- and u-channels are the following, 






\begin{figure}[t]
\centerline{%
\includegraphics[width=10.5cm]{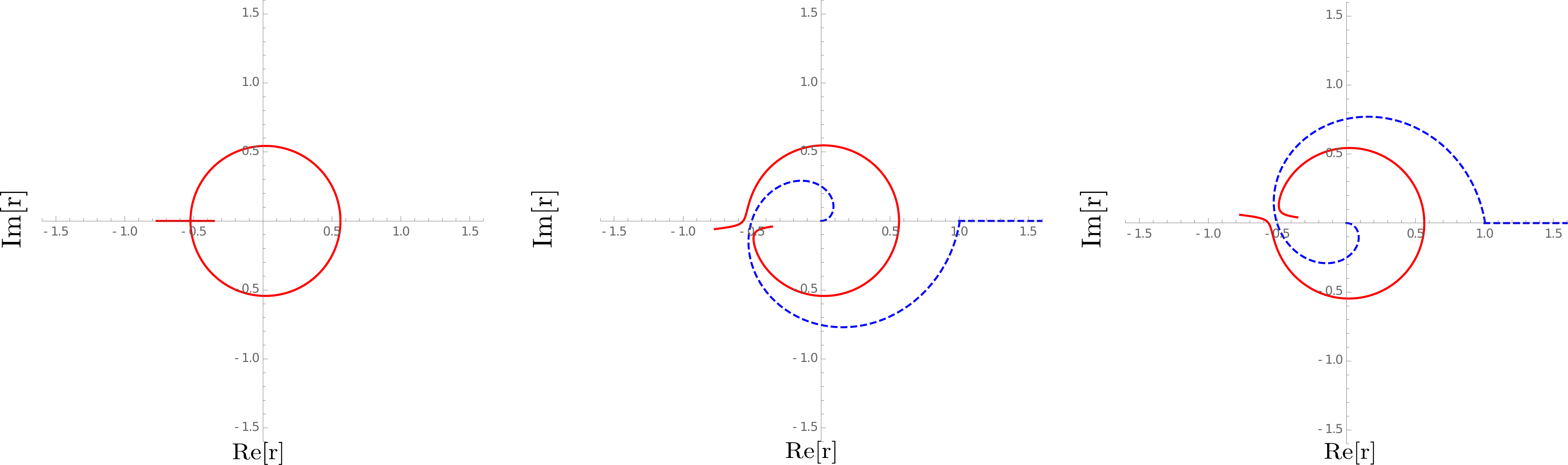}}
\caption{The solid line corresponds to the branch cuts due to the two pion propagators. At the threshold there is not path deformation possible (left). We add a negative (middle) or positive (right) imaginary part to $P^2$ to get an opening in the cuts. The dotted line shows a possible integration path. }
\label{Fig:cuts}
\end{figure}

\scriptsize
\begin{eqnarray}
K^{(s)}_{da,he}(q,p.r;P)&=& \Bigg[\frac{1}{2} [\Gamma_{\pi}^j]_{dc} \left(p + \frac{P}{4} - \frac{r}{4}; \frac{P + r}{2} \right) S_{cb}\left(p - \frac{r}{2}\right)[\Gamma_{\pi}^j]_{ba} \left(p - \frac{P}{4} - \frac{r}{4}; \frac{P - r}{2} \right)  \nonumber \\ 
 &\times & \frac{1}{2} [\Gamma_{\pi}^j]_{hg} \left(q + \frac{P}{4} - \frac{r}{4}; \frac{r - P}{2} \right) S_{gf}\left(q - \frac{r}{2}\right)[\Gamma_{\pi}^j]_{fe} \left(q + \frac{P}{4} - \frac{r}{4}; -\frac{P + r}{2} \right) \nonumber \\&\times& D_{\pi}\left(\frac{P + r}{2}\right) D_{\pi}\left(\frac{P - r}{2}\right) ~,
\label{newkernel}
\end{eqnarray}

\begin{eqnarray}
K^{(u)}_{da,he}(q,p.r;P)&=& \Bigg[\frac{1}{2} [\Gamma_{\pi}^j]_{dc} \left(p + \frac{P}{4} + \frac{\tilde{r}}{4}; \frac{P - \tilde{r}}{2} \right) S_{cb}\left(p + \frac{\tilde{r}}{2}\right)[\Gamma_{\pi}^j]_{ba} \left(p - \frac{P}{4} + \frac{\tilde{r}}{4}; \frac{P + \tilde{r}}{2} \right)  \nonumber \\ 
 &\times & \frac{1}{2} [\Gamma_{\pi}^j]_{hg} \left(q + \frac{P}{4} - \frac{\tilde{r}}{4}; \frac{\tilde{r} - P}{2} \right) S_{gf}\left(q - \frac{\tilde{r}}{2}\right)[\Gamma_{\pi}^j]_{fe} \left(q + \frac{P}{4} - \frac{\tilde{r}}{4}; -\frac{P + \tilde{r}}{2} \right) \nonumber\\  &\times& D_{\pi}\left(\frac{P + \tilde{r}}{2}\right) D_{\pi}\left(\frac{P - \tilde{r}}{2}\right) ~.
\label{newkernel1}
\end{eqnarray}
\normalsize


\subsection{Branch-cut structure}

The inclusion of the two kernels given in equations (\ref{newkernel}) and (\ref{newkernel1}) in BSE calculations is very challenging, as they have a non-trivial analytic structure. Knowing the position of the singularities allows to develop effective algorithms for numerical calculations (see, e.g. \cite{Weil:2017knt}). For example, the kernel features now branch cuts corresponding to the virtual pions. Those are determined by the zeroes of the denominators and are parametrized by (see Fig.~\ref{Fig:cuts})
\begin{eqnarray}
y_1(P,zr) &=& -m_{\pi}^2 - P^2 + 2 z_r^2 P^2 - 2\sqrt{-m_{\pi}^2 z_r^2 P^2 - z_r^2 P^4 +  z_r^4 P^4} \nonumber\\
y_2(P,zr) &=& -m_{\pi}^2 - P^2 + 2 z_r^2 P^2 + 2\sqrt{-m_{\pi}^2 z_r^2 P^2 - z_r^2 P^4 +  z_r^4 P^4} 
\end{eqnarray}


\begin{figure}[t]
\centerline{%
\includegraphics[width=8.5cm]{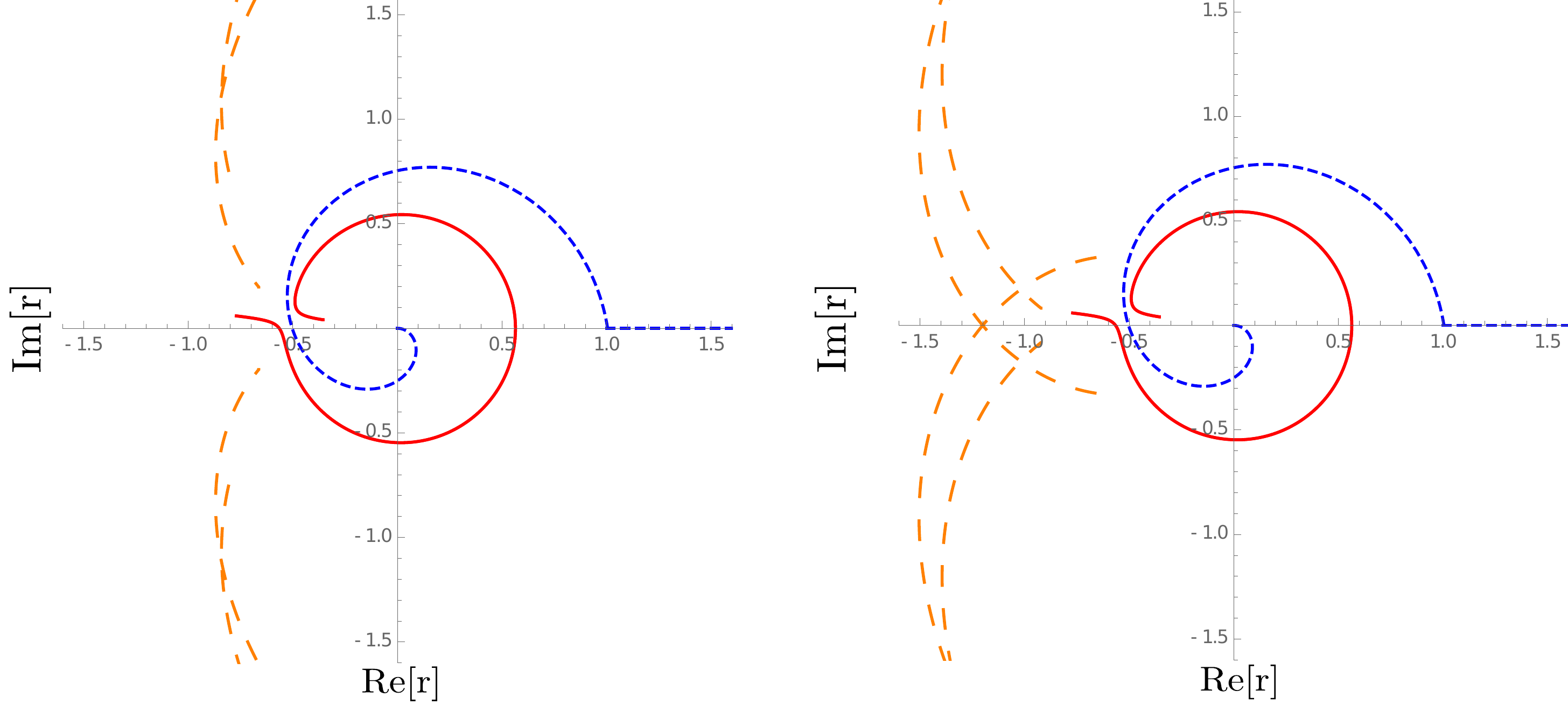}}
\caption{Branch cuts due to the quark propagators (dashed lines), for two different values in the relative momentum $p$. }
\label{Fig:cutspoles}
\end{figure}

\noindent In order to perform the integration to solve the BSE with the new contributions, first we need to deform the contour since the branch cut overlaps the real axis. In Figure \ref{Fig:cuts} we sketch a possible integration path avoiding the cut. Additionally the quark propagators contribute with more cuts. One way to parametrize the quark propagator $S(p) = -i \slashed{p} \sigma_v(p^2) + \sigma_s(p^2)$ is as a sum of complex-conjugate poles (see e.g. \cite{El-Bennich:2016qmb})
\begin{equation}
\sigma_v = \sum_{i}^n \left[\frac{\alpha_i}{p^2 + m_i} + \frac{\alpha_i^\ast}{p^2 + m_i^\ast}\right], \quad \sigma_s = \sum_{i}^n \left[\frac{\beta_i}{p^2 + m_i} + \frac{\beta_i^\ast}{p^2 + m_i^\ast}\right] ,
\end{equation}
\noindent where the parameters $m_i$, $\alpha_i$, $\beta_i$ can be obtained by fitting the corresponding solution along the real axis of $p^2$. In Figure \ref{Fig:cutspoles} we plot the additional cuts from the quark propagator. Nevertheless, these cuts do not cross the real axis, and do not affect the integration contour.

\section{Summary and Outlook}
In this work we observed that in order to include the resonant character of bound states in BSE calculations, virtual decay mechanism must be included. Moreover, we studied the resulting analytic structure. The appearance of branch cuts entails that the integration contour must be deformed in order to avoid the crossing of the cuts. The quark propagators contribute with additional cuts but these do not overlap the integration path.
\section{Acknowledgments}
This work was supported by the project P29216-N36 and the Doctoral Program W1203-N16 “Hadrons in Vacuum, Nuclei and
Stars”, both from the Austrian Science Fund, FWF. 

\small
\bibliographystyle{polonica}
\bibliography{mybib}{}

\end{document}